\documentclass{elsart}
\usepackage{graphics}
\usepackage{psfrag}
\usepackage{epsfig}
\usepackage{epsf}
\usepackage{float}
\usepackage{amssymb,stmaryrd,latexsym}
\usepackage{amscd}
\usepackage{mathrsfs}
\newcommand{\ea}{{\textit{et al.}}}
\newcommand{\be}{\begin{eqnarray}}
\newcommand{\ee}{\end{eqnarray}}
\newcommand{\beq}{\begin{equation}}
\newcommand{\eeq}{\end{equation}}

\usepackage{amsmath}
\usepackage{amssymb}
\usepackage{epsfig}

\newcommand{\dis}{\displaystyle}

\newcommand{\mpi}{M_{\pi}}
\newcommand{\fpi}{F_{\pi}}
\newcommand{\boldpi}{\mbox{\boldmath $\pi$}}

\newcommand{\boldtau}{\mbox{\boldmath $\tau$}}

\begin{document}
\hfill{\small FZJ--IKP(TH)--2009--25, HISKP-TH-09-28}

\begin{frontmatter}
\title{Extraction of the strong neutron-proton mass difference from the
charge symmetry breaking in $pn\to d\pi^0$}

\author{A. Filin$^{1,2}$, V. Baru$^{1,2}$, E.~Epelbaum$^{1,3}$, J.~Haidenbauer$^{1,4}$, }
\author{ C.~Hanhart$^{1,4}$, A. Kudryavtsev$^2$, and U.-G. Mei\ss ner$^{1,3,4}$}

{\small $^1$ Institut f\"{u}r Kernphysik  (Theorie) and J\"ulich Center for Hadron Physics,}\\
{\small Forschungszentrum J\"ulich,  D--52425 J\"{u}lich, Germany} \\
{\small $^2$ Institute for Theoretical and Experimental Physics,} \\
{\small 117218, B. Cheremushkinskaya 25, Moscow, Russia} \\
{\small $^3$ Helmholtz-Institut f\"{u}r Strahlen- und Kernphysik (Theorie) and 
} \\ 
{\small Bethe Center for Theoretical Physics, Universit\"at Bonn, D-53115 Bonn, Germany}\\
{\small $^4$  Institute for Advanced Simulation,}\\
{\small 
Forschungszentrum J\"ulich,  D--52425 J\"{u}lich, Germany}\\

\begin{abstract}
\noindent 
We perform a complete calculation of charge symmetry breaking effects 
for the reaction $pn\to d\pi^0$ at leading order in
chiral perturbation theory. A new leading- order operator is included.
From our analysis we extract $\delta m_N^{\rm str}$, the strong contribution to
the neutron--proton mass difference. The value obtained, 
$\delta m_N^{\rm str} = (1.5 \pm 0.8 \ {\rm (exp.)} \pm 0.5 \ {\rm (th.))} 
\ {\rm MeV}$, 
is consistent with the result based on the Cottingham sum rule. 
This agreement provides a non--trivial
test of our current understanding of the chiral structure of QCD.
\end{abstract}

\end{frontmatter}

{\bf 1.} At the fundamental level of the Standard Model, isospin violation
is due to quark mass differences as well as electromagnetic
effects~\cite{Wein,Gasser,Leut}. Amongst the isospin violating
effects in hadronic reactions the ones that are charge symmetry breaking (CSB),
i.e. that emerge from an interchange of up and down quarks,
are of particular interest. Their importance is due to the fact that the
neutral--to--charged pion mass difference, which is  almost entirely of
electromagnetic origin and usually dominates isospin violating hadronic
observables, does not contribute here. Therefore, the
sensitivity to the quark mass difference $m_d - m_u$ is more pronounced 
in observables related to CSB.

CSB effects manifest themselves in many different physical phenomena
such as  the mass splitting of hadronic isospin multiplets
(e.g.~$ m_n\ne m_p$~\cite{Gasser} and $M_{D^0}\ne M_{D^+}$~\cite{ourD}),
$\eta$--decays (for a recent two-loop calculation, see~\cite{Hans} and
references therein), the
different scattering lengths of $nn$ and $pp$ systems after removing
electromagnetic effects in $pp$ scattering (see, e.g.~the review
article \cite{MOS}), 
neutron-proton elastic scattering at
intermediate energies \cite{NNelast}, hadronic mixing 
(e.g.~$\rho^0-\omega$ \cite{Barkov} or $\pi^0-\eta$ \cite{coon} mixing) and
the binding-energy difference of mirror nuclei known as Nolen-Schiffer
anomaly \cite{Nolen}.
Recently, experimental evidence for CSB was found in reactions
involving the production of neutral pions. At IUCF non-zero values for the
$dd\to \alpha \pi^0$ cross section were established \cite{Stephenson}. 
At TRIUMF a forward-backward
asymmetry of the differential cross section for $pn\to d\pi^0$
was reported which amounts to $A_{fb}=[17.2
\pm 8 {\rm (stat.)} \pm 5.5 {\rm (sys.)}] \times 10^{-4}$~\cite{Opper}.
In a charge symmetric world the initial $pn$ pair would 
consist of identical nucleons in a pure isospin one state and thus an 
interchange of beam and target would have no observable impact so that
the cross section should be symmetric. Thus, the apparent forward--backward
asymmetry is due to charge symmetry breaking.

A solid theoretical background for investigating CSB effects is 
provided by chiral
perturbation theory (ChPT), the low-energy effective field theory of
QCD \cite{chpt1,chpt2,ulfbible}. Especially, since electromagnetic
and quark mass (strong) effects typically contribute with similar
strength,
they can only be disentangled within a systematic effective field
theory.  ChPT has been recently extended to pion production
reactions, i.e. to processes with a large initial momentum $p\simeq
\sqrt{m_N \, M_\pi} \simeq 360\,$MeV, with $\mpi (m_N)$  the 
pion (nucleon) mass. The proper way to include this scale in the 
power counting was presented in Ref.~\cite{ch3body} and implemented in
Ref.~\cite{withnorbert}, see Ref.~\cite{report} for a review
article. Within this scheme it turned out to be possible to achieve a
quite good theoretical description of $s$-wave pion production in $pp\to
d\pi^+$ at next-to-leading (NLO) order \cite{lensky2}; $p$-wave pion
production in different channels of $NN\to NN\pi$ at
next-to-next-to-leading (N$^2$LO) order was investigated in
Ref.~\cite{pwave}. These developments in our
understanding of isospin conserving pion production mechanisms provide
a very good starting point for studying isospin violation effects in
$pn\to d\pi^0$ and $dd\to\alpha \pi^0$. First efforts in this
direction were already presented in Refs.~\cite{kolck,bolton,nisk} for 
the $pn\to d\pi^0$ reaction and in Refs.~\cite{gard,nogga,lahde,fonseca}
for $dd\to\alpha \pi^0$. 
In this work we improve the theory for the former reaction.

\begin{figure}[tb]
\begin{center}
\includegraphics[scale=0.9,clip=]{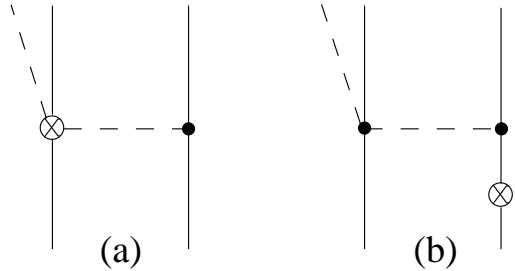}
\end{center}
%\vspace*{8pt}
\caption{\label{diagLO}Leading order diagrams for the isospin
  violating $s$-wave amplitudes of $pn\to d\pi^0$.
%Isospin violating operators are denoted by $\otimes$.
 Solid (dashed) lines denote
 nucleons (pions). Diagram (a) corresponds to isospin
 violation in the $\pi N$ scattering vertex explicitly whereas  diagram (b)
 indicates an isospin-violating contribution due to the neutron--proton 
 mass difference in conjunction with the time-dependent
  Weinberg-Tomozawa operator (see text for details).  }
\end{figure}

The neutron--proton mass difference is due to strong and electromagnetic
interactions \cite{Gasser}, i.e. 
$\delta m_N=m_n-m_p=\delta m_N^{\rm str}+\delta m_N^{\rm em}$.
As a result of the chiral structure of the QCD Lagrangian, the
strength of the rescattering operator  in $pn\to d\pi^0$ depicted in Fig.~\ref{diagLO}(a)
is proportional to a different combination of $\delta m_N^{\rm str}$ and 
$\delta m_N^{\rm em}$  \cite{kolck,ulfsven} (for related work on isospin violation 
in pion-nucleon scattering  see \cite{Fettes:2000vm}). 
Thus, the analysis of CSB effects in $pn\to d\pi^0$ should allow
to determine the values of $\delta m_N^{\rm str}$ and 
$\delta m_N^{\rm em}$ individually.
This was for the first time stressed and exploited 
in Ref.~\cite{kolck}.
Consistency of these important quantities as determined from
$pn\to d\pi^0$, where they control the strength of the isospin
violating $\pi N$ scattering amplitude, with results obtained
from the neutron--proton mass difference itself~\cite{Gasser} 
employing the Cottingham sum rule \cite{Cottingham},
would provide a highly non--trivial test of our current 
understanding of QCD. 
It was therefore quite disturbing to find that, using
the values for $\delta m_N^{\rm str}$ and $\delta
m_N^{\rm em}$ from Ref.~\cite{Gasser}, the leading order calculation
of the forward-backward asymmetry \cite{kolck} over-predicted the
experimental value by about a factor of 3 --- a consistent description
would call for an agreement with data within the theoretical
uncertainty of 15\% for this kind of calculation\footnote{It was shown
  in Ref.~\cite{gard} that there is no NLO contribution -- thus the
  theoretical uncertainty of a leading order calculation is expected
  to be of the order of $M_\pi/m_N$.}. The evaluation of certain
higher order corrections performed in Ref.~\cite{kolck} and in a very
recent study~\cite{bolton} did not change the situation noticeably
--- the significant overestimation of the data persisted.

In this Letter we show that there is one more rescattering
operator that contributes at LO. We evaluate this new LO operator 
and we also recalculate the LO contribution considered in 
Ref.~\cite{kolck} since the numerical evaluation in that work turned 
out to be incorrect \cite{Niskanen}.
The complete LO calculation for
$pn\to d\pi^0$ reveals a very good agreement with the experimental
data. Moreover, the resulting contribution is found to be
proportional to $\delta m_N^{\rm str}$ only. Thus, a quantitative 
understanding of the CSB part of $pn\to d\pi^0$ promises an alternative 
method of extraction of this important quantity compared to that used in
Ref.~\cite{Gasser}. 

% \section{General remarks}
 
{\bf 2.} The differential cross section of the reaction $pn\to d\pi^0$ can be
 expanded into a series of Legendre polynomials $P_i(\cos\theta)$.  In the near-threshold 
region only the first terms are relevant
\begin{equation}
\label{eq_DXS_expansion}
\frac{d \sigma}{d \Omega} (\theta) = A_0 + A_1 P_1(\cos\theta) + \cdots ,
\end{equation}
where $\theta$ is the angle between the incident proton and the pion produced
and the $A_i$ are functions depending on the different
partial wave amplitudes.
Due to CSB effects the differential cross
section %$\frac{d \sigma}{d \Omega} (\theta) \neq \frac{d \sigma}{d \Omega} (\pi - \theta)$ i.e. it
is not symmetric with respect to the replacement $\theta
\leftrightarrow \pi - \theta$ and thus $A_1$ is non--vanishing.  The
forward-backward asymmetry is defined as
\begin{equation}
\label{Afb_definition_CS}
A_{fb} = \frac{ \int\limits_0^{\pi/2} \left[ \frac{d \sigma}{d \Omega} (\theta) - \frac{d \sigma}{d \Omega} (\pi - \theta)\right] \sin\!\theta d\theta}
              { \int\limits_0^{\pi/2} \left[ \frac{d \sigma}{d \Omega} (\theta) + \frac{d \sigma}{d \Omega} (\pi - \theta)\right] \sin\!\theta d\theta}
 = \frac{A_1}{2 A_0} \ ,
\end{equation}
where we used Eq.~(\ref{eq_DXS_expansion}) in the last equality.  The
experiment at TRIUMF was done very close to threshold at
$T_{{\rm lab}}=279.5$~MeV, which is equivalent to an excess energy of about
2 MeV or $\eta=0.17$ --- traditionally, the energy for pion production
reactions is given in terms of $\eta$, the pion momentum in units of
the pion mass. At this energy the total cross section $\sigma=4\pi
A_0$ is dominated by the isospin conserving $s$-wave pion production
amplitude.  At present, this quantity is known theoretically only 
up-to-and-including terms at NLO which implies a theoretical uncertainty
of the order of 30\% for the cross section \cite{lensky2}.  Therefore, to minimize
the uncertainty of the current study, we use the experimental value
for $\sigma (nn\to d\pi^-)=252^{+5}_{-11}\ \cdot \eta \ [\mu$b] extracted with
very high accuracy from the lifetime of the pionic deuterium 
atom\footnote{Note that
 the Coulomb corrections were already removed in the extraction of
 this quantity from pionic atoms, see, e.g., the review~\cite{akaki}.},
measured at PSI \cite{pidexp}. 
To convert this number to the reaction of interest here
we may
use isospin symmetry which gives $\sigma(pn\to d\pi^0)=\sigma(nn\to d\pi^-)/2$.
Isospin violating 
effects in this relation are to be expected of natural size and thus 
will not further be considered.
In addition, we include in $A_0$ also the contribution from the $p$--wave
 production. Here we take the results of the N$^2$LO calculation
of Ref.~\cite{pwave}. Thus, we get in total
$A_0=10.0^{+0.2}_{-0.4} \cdot \eta  \ + (47.8\pm 5.7)\cdot \eta^3 \  [\mu$b]. 

At the energies we consider here, the function $A_1$ depends on
the interference of either an isospin conserving (IC) $p$-wave and an
isospin violating (IV) $s$-wave amplitude or of an IV $p$-wave with an IC
$s$-wave. However, only the former piece contributes at leading order.
Thus, to the order we are working, one can write
\begin{equation}
\label{eq_A_one}
A_1 = \frac{1}{128\pi^2} \frac{\eta M_\pi}{p(\mpi + m_d)^2} %\frac{1}{4}
%\Bigg[
\, {\rm Re}\left[\left(M^{\rm IC,p}_1+ \frac23 M^{\rm IC,p}_2\right) M^{\rm IV,s^*}\right]
%\Bigg]
\end{equation}
where $m_d$ is the deuteron mass and $k_{\pi}$  the pion momentum.
Here, $M^{\rm IC,p}_1$ and $M^{\rm IC,p}_2$ are the invariant
amplitudes corresponding to the isospin conserving $p$-wave pion
production in the $^1S_0\to {}^3S_1p$ and $^1D_2\to {}^3S_1p$ partial
waves and $M^{\rm IV,s}$ is the corresponding amplitude for the
isospin violating $s$-wave production in the $^1P_1\to {}^3S_1s$ partial
wave. Thus, in the latter amplitude the isovector pion is produced 
from an isoscalar $NN$ pair
($I_i=0$). In this work we use the IC $p$-wave amplitudes of
Ref.~\cite{pwave}. 
As explained in this reference, the 
contribution $M^{\rm IC,p}_1$ is quite uncertain and negligibly small. We
therefore negelect its contribution in this calculation.
 The IV $s$-wave amplitude is discussed in detail
below.

{\bf 3.} Our calculations are based on the effective chiral Lagrangian
\cite{OvK,ulfbible} which reads
\begin{equation}
 \mathcal{ L}^{(0)}  = 
% \nonumber
   N^{\dagger}\left[\frac{1}{4 F_{\pi}^{2}} \boldtau \cdot
         (\dot{\boldpi}\times{\boldpi})
         +\frac{g_{A}}{2 F_{\pi}}
         \boldtau\cdot\vec{\sigma}\cdot\vec{\nabla}\boldpi
\right]N 
%+\frac{h_{A}}{2 F_{\pi}}\left[N^{\dagger}(\boldT\cdot
%          \vec{S}\cdot\vec{\nabla}\boldpi)\Psi_\Delta +h.c.\right] 
+\cdots \ \ ,  
\label{la0}
\end{equation} 
for the leading $\pi N$ interaction terms relevant for our study.
The leading isospin-violating terms, generated by the quark--mass
difference and hard-photon contributions, are 
\beq
\mathcal{ L}^{(0)}_{\rm iv} =  \frac{\delta m_{N}}{2}
\; N^{\dagger}\tau_3 N 
-\frac{\delta m_{N}^{\rm str}}{4 F_{\pi}^2} 
\; N^{\dagger}\boldtau\cdot\boldpi \pi_3 N 
-\frac{\delta m_{N}^{\rm em}}{4 F_{\pi}^2} 
\; N^{\dagger}(\tau_3\boldpi^2-\boldtau\cdot\boldpi\pi_3 ) N 
+\ldots 
\label{liv}
\eeq
with $\delta m_N=\delta m_N^{\rm str}+\delta m_N^{\rm em}$.
The ellipses stand for further terms which are not relevant here. 
In the equations above $F_\pi$ denotes the pion decay constant in the chiral limit,
 $g_A$ is the axial-vector coupling of the nucleon and $N$ ($\boldpi$) corresponds to 
the nucleon (pion) field. More precisely, this form of the IV strong and electromagnetic 
operators is only correct at leading order and neglecting terms with more than
two pion fields. The more generic form involves the low-energy constants (LECs) $c_5$
and $f_2$ (for precise definitions, see e.g.~\cite{ulfsven}). Also, beyond LO 
other strong and electromagnetic LECs will have to be taken into account.

The diagrams that contribute to the amplitude $M^{\rm IV,s}$ at LO are
shown in Fig.~\ref{diagLO}. Diagram (a) corresponds to the
rescattering process in which CSB occurs explicitly in the $\pi N$
scattering
vertex due to the last two terms in Eq.~(\ref{liv}). In diagram (b)
pion rescattering proceeds via
the Weinberg-Tomozawa operator (first term in Eq.~(\ref{la0}))
which produces an additional isospin violating piece from
the mass difference of neutron and proton due to its
time dependence as will be discussed later in this section.

In order to understand the interplay of diagram~(a) and 
diagram~(b) of Fig.~\ref{diagLO} it is sufficient
to focus on the $\pi N$ rescattering vertex on nucleon~1.
From the 
pion production vertex on nucleon~2 we only keep the isospin structure,
for the rest is identical for both diagrams.
  The relevant part of diagram (a) then reads 
\beq
\hat I_{\rm (a)}= -i\frac{ \delta m_N^{\rm str}}{4\fpi^2} \left( \boldtau^{(1)} \cdot
  \boldtau^{(2)} + \tau^{(1)}_3 \tau^{(2)}_3\right) +i\frac{
  \delta m_N^{\rm em}}{4\fpi^2} \left( \boldtau^{(1)} \cdot \boldtau^{(2)}
  -\tau^{(1)}_3 \tau^{(2)}_3\right).  
\label{Iaop}
\eeq 
We work at leading order in IV. Since we
study an IV transition operator, we may therefore treat the
external nucleons as identical particles --- this is not the case for the 
diagram~(b), where the mass difference of the external particles plays 
the essential role.
%Due to the fact that isospin violation in the diagram (a) 
%occurs in the $\pi N$ vertex explicitly nucleons in the evaluation of the 
%isospin matrix elements with this operator can be treated as identical particles. 
The evaluation of the operator Eq.~(\ref{Iaop}) for the isospin violating transition
from the isospin zero initial $pn$ state to the isospin zero deuteron
state yields 
\beq \langle I_f=0|\hat I_{\rm (a)}|I_i=0\rangle = \frac{ i}{4\fpi^2}\ 4\left
  (\delta m_N^{\rm str} -\delta m_N^{\rm em}/2 \right ) \ . 
\label{Ia}
\eeq 
This piece
represents the complete rescattering contribution included in
Refs.~\cite{kolck,bolton}.
\begin{figure}[tb]
\begin{center}
\includegraphics[scale=0.9,clip=]{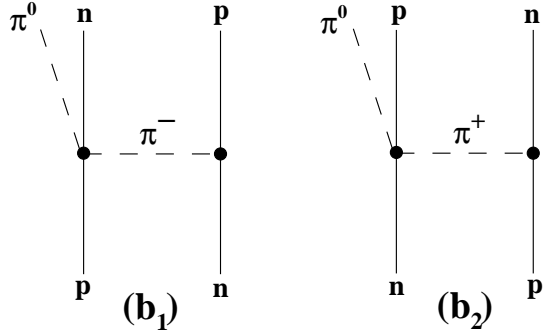}
\end{center}
%\vspace*{8pt}
\caption{\label{diagWT} Leading-order contributions to isospin violation due to 
the time-dependent Weinberg-Tomozawa operator in the particle basis.
}
\end{figure}
Let us now look more closely at diagram~(b) of Fig.~\ref{diagLO}.  The relevant part of
the amplitude for this diagram can be most easily calculated in the
particle basis as shown in Fig.~\ref{diagWT}. One gets \beq \langle I_f=0|\hat
I_{(\rm b)}|I_i=0\rangle = -\frac{1}{2} (I_{\rm b_1}+I_{\rm b_2}),
\label{Ib}
\eeq where $I_{\rm b_1}$ and $I_{\rm b_2}$ are the isospin coefficients
corresponding to the diagrams (b$_1$) and (b$_2$) of Fig.~\ref{diagWT} and the
factor $-1/2$ stems from the Clebsch-Gordan coefficients.  Note that, since
the WT operator involves a time derivative, the corresponding Feynman rule reads
\beq 
V^{ab}_{WT}=\frac{1}{4\fpi^2}
\varepsilon_{abc}\tau_c (q_0 + \mpi) \ , 
\eeq
 with $a,b$ and $c$ Cartesian pion indices and $q_\mu$ the four-momentum of the
intermediate pion.
Due to the explicit appearance of $q_0$ in $V_{WT}$, the final expression for 
diagram $\rm (b)$ of Fig.~\ref{diagLO} depends on the
neutron--proton mass difference. Indeed, the evaluation of this
vertex for the diagrams (b$_1$) and $\rm (b_2)$ of Fig.~\ref{diagWT} yields 
\be \displaystyle %\vartheta
&&{V_{WT}\ = \ } \frac{-i}{4\fpi^2}\left\{
\begin{array}{rr}
 {\sqrt{2}\left(\frac{\dis 3\mpi}{\dis 2}+\delta m_N \right)} &   {\rm for\ \ diagram}\ \ (b_1),\\
{-\sqrt{2}\left(\frac{\dis 3\mpi}{\dis 2}-\delta m_N \right)} & \ \ \ \ \  {\rm for\ \ diagram}\ \ (b_2).
%{\Gamma_{1s}}=8\alpha^3\mu_r^2 & (b_2)\\ 
\end{array}
\right.
\ee
Thus, in the isospin violating contribution to Eq.~(\ref{Ib}) the terms 
$\propto M_\pi$  cancel while those $\propto \delta m_N$ survive.
The non-vanishing isospin matrix element
for the diagram~(b) of Fig.~\ref{diagLO} amounts to 
 \beq \langle I_f=0|\hat I_{(\rm b)}|I_i=0\rangle =
\frac{i}{4\fpi^2}\ 2\delta m_N \ .
\label{Ibfinal}
\eeq 
Adding up the contributions of diagrams~(a) and (b) we find that the
resulting contribution at LO depends on the quark mass contribution to
the nucleon mass difference only --- the electromagnetic piece
vanishes completely:
 \beq \langle I_f=0|\hat I_{(\rm a)}+\hat I_{(\rm b)}
|I_i=0\rangle = \frac{i}{4\fpi^2}\ 6\, \delta m_N^{\rm str}.
\label{sum}
\eeq 
In comparison with
the expression used previously (cf. Eq.~(\ref{Ia})) the
rescattering operator gets enhanced by about 30\%, when
standard values  $\delta m_N^{\rm str}=2$ MeV and $\delta m_N^{\rm em}=-0.76$ MeV~\cite{Gasser} are used.

An alternative method to derive the same result is by using the
field-redefined Lagrangian as discussed in
Refs~\cite{friar,friar2,Epelbaum:2007sq} 
--- see also Ref.~\cite{Epelbaum:2004xf} where unitary
transformations are used. In this formulation the pion and nucleon
fields are redefined in order to eliminate the first term in the
effective Lagrangian in Eq.~(\ref{liv}).  This allows one to work with
nucleons as indistinguishable particles. All terms in the Lagrangian
are invariant under this transformation except the ones involving a
time derivative such as the Weinberg-Tomozawa operator which generates
an additional isospin violating $\pi N\to \pi N$ vertex $\propto
\delta m_N$ that cancels exactly the electromagnetic contribution to
this vertex $\propto \delta m_N^{\rm em}$. 

It should be stressed that also in Ref.~\cite{kolck} some effects
from the neutron--proton mass difference were included, using
the formalism of Ref.~\cite{nisk}. However, these effects appear
effectively in the isospin violating $\pi NN$ vertex and are explicitly 
in conflict with the chiral structure of QCD. Therefore, they are very 
different from those discussed above.

For the sake of completeness, we present here the tree-level invariant amplitude
$M^{\rm IV,s}_{\rm tree}$ corresponding to the LO calculation 
\beq
M^{\rm IV,s}_{\rm tree}=-i\frac{12m_N^2g_A}{\fpi^3}\, 
\delta m_N^{\rm str}\, %(   {\vec {\mathcal{S}}}\,' \hat p) \mathcal{I}
\ \,\int \frac{d\Omega_{p\,'}}{4\pi} \frac{(\vec p\,'-\vec p\,)\cdot \hat p}
{(\vec p\,'-{\vec p}\,)^2+\mpi^2},
\eeq
where $\vec p$ and $\vec p\,'$ denote 
initial and final relative momenta of the two nucleons, respectively, 
and $\hat p=\vec p/p$. In the calculation we use $\fpi=92.4$ MeV 
and $g_A=1.32$ (utilizing the Goldberger-Treiman relation). 
To get the full amplitude $M^{\rm IV,s}$ which enters the observables,
$M^{\rm IV,s}_{\rm tree}$ given above needs to be convoluted with
proper $NN$ wave functions in the initial and final states, 
cf.~Appendix A of Ref.~\cite{pwave} for a detailed description. 
Ideally, one should use wave
functions derived in the same framework, namely ChPT. However, up to
now these are only available for energies below the pion production 
threshold~\cite{NN}. We therefore adopt 
the so-called hybrid approach, first introduced by Weinberg~\cite{Weinberg:1991um}, 
i.e. we use transition operators derived within effective field theory and
convolute them with realistic $NN$ wave functions~\cite{CCF}. 

Now we are in the position to discuss the results for the
forward-backward asymmetry within the complete LO calculation.  
Using the values for the parameters specified above and utilizing the $NN$ wave
functions from Ref.~\cite{CCF}, the
result can be presented in the form
 \beq 
A_{\rm fb}^{\rm LO} = (11.5 \pm 3.5)\times 10^{-4} \ 
\frac{\delta m_N^{\rm str}}{\rm MeV} \ .
\label{AfbLO}
\eeq
As discussed above, the calculation of the
coefficient has a theoretical uncertainty of 15\% which is
doubled to provide a more conservative estimate.
This uncertainty is included in the expression above.
We now use the experimental result for $A_{\rm fb}$ \cite{Opper} to extract
$\delta m_N^{\rm str}$ which yields  
\beq
\delta m_N^{\rm str} = 
\left(1.5 \pm 0.8 \ {\rm (exp.)} \pm 0.5 \ {\rm (th.)}\right) \ {\rm MeV} \ ,
\eeq
where we added the experimental errors in quadrature.
This is the final result of our analysis. 
At the present stage, the uncertainty in the determination of $\delta m_N^{\rm
str}$ is dominated by the experimental uncertainty for $A_{\rm fb}$.

In this context
let us point out the following: Besides the additional IV contribution
discussed in detail above there are other reasons why our result
deviates from those of Refs.~\cite{kolck,bolton} already at leading
order.  The numerical evaluation of the diagram (a) of
Fig.~\ref{diagLO} revealed that the value we obtain is significantly
smaller than the one found in Ref.~\cite{kolck}. It turned out that
the result of that work is too large by a factor of 4 due to an error
\cite{Niskanen}.  The discrepancy of our result to that of
Ref.~\cite{bolton} is an accumulation of various effects. First of all
in Ref.~\cite{bolton} the isospin conserving $s$-- and $p$--wave
amplitudes are calculated within ChPT up to NLO. Thus, they come with
individual uncertainties of 30 \% and 15 \%, respectively --- the
uncertainty for the $s$--wave appears doubled for this amplitude,
since it enters squared in $A_0$, while the $p$--wave amplitudes
mainly contribute linearly to $A_1$ ---
cf. Eqs.~(\ref{Afb_definition_CS}) and (\ref{eq_A_one}). In contrast to this we take the
$s$--wave amplitude directly from data, with a negligible uncertainty
and for the $p$--wave amplitudes the results of Ref.~\cite{pwave},
which were calculated to NNLO and are additionally constrained by
data. Thus, combining these uncertainties with that for the CSB
amplitude in quadrature, a total uncertainty of 50 \% arises for the
result of Ref.~\cite{bolton}. In addition, the $p$--wave
amplitude with the $^1S_0$ initial state employed in Ref.~\cite{bolton}, which amounts to an
enhancement of 50\% in the isospin conserving $p$--wave
amplitude in this calculation, is in conflict with the data
for $pp\to d\pi^+$, which calls for a negligible contribution
of this partial wave~\cite{pwave}. 
These effects together --- the larger uncertainty
of the calculation of Ref.~\cite{bolton} as well
as the wrong $p$--wave amplitude --- explain the discrepancy
between our result and that of Ref.~\cite{bolton}.

In Ref.~\cite{kolck} also some higher order contributions
were calculated, see also  \cite{MOS}.
While individually sizeable, the sum of the considered corrections 
was found to contribute very little to the asymmetry. We re-evaluated these 
additional pieces and confirmed these findings qualitatively though our
results deviate from the ones of Refs.~\cite{kolck,MOS} quantitatively~\cite{filin}.
In addition, in Ref.~\cite{bolton} some CSB $p$--wave amplitudes were 
evaluated. Through an interference with the isospin conserving $s$--wave
they also contribute to the forward--backward asymmetry discussed
in this work, however, only at NNLO. It is reassuring that quantitatively
these contributions are in line with the power counting estimates
given above and thus support our uncertainty estimate.

{\bf 4.} In this work we calculated the CSB forward--backward
asymmetry for the reaction $pn\to d\pi^0$ to leading order in the
chiral expansion. We showed that the resulting production operator is driven
by that contribution to the neutron-proton mass difference which is coming
solely from the quark mass difference, $\delta m_N^{\rm str}$. 
Using the TRIUMF measurement of the
forward-backward asymmetry \cite{Opper} we extracted
\beq
\delta m_N^{\rm str} = 1.5 \pm 0.9  \ {\rm MeV} \ ,
\label{dmpiprod}
\eeq
where the theoretical and experimental uncertainties are added in quadrature.
This number is to be compared with the value for
the same quantity extracted from the neutron--proton mass
difference --- employing the Cottingham sum rule~\cite{Cottingham}
to determine the electromagnetic contribution to the 
mass difference to $\delta m_N^{\rm em} = -0.76 \pm 0.3  \ {\rm MeV}$~\cite{Gasser} ---
\beq
\delta m_N^{\rm str} = 2.0 \pm 0.3  \ {\rm MeV} \ .
\label{dmmn}
\eeq
This value is consistent with a recent determination
of the same quantity using lattice QCD~\cite{lattice},
$\delta m_N^{\rm str}=2.26 \pm 0.57 \pm 0.42 \pm 0.10$ MeV
was found, where the uncertainties emerge from statistics,
from the input as well as from the chiral extrapolation. 
We emphasize that the agreement of the various independent 
extractions provides a highly non-trivial and important test 
for our understanding of the chiral symmetry and the 
isospin breaking pattern of QCD,
since Eq.~(\ref{dmpiprod}) is obtained from a reaction
where $\delta m_N^{\rm str}$ is governed by the strength of $\pi N$ scattering,
while Eq.~(\ref{dmmn}) is derived from the neutron--proton mass
difference itself. 
The link between these two apparently very different
physical quantities is provided by the symmetry pattern of QCD
properly implemented in hadronic matrix elements through chiral perturbation theory.

At present the uncertainty in Eq.~(\ref{dmpiprod}) is dominated by the
experimental error bars -- an improvement on this side would be very
important. Still, a more refined calculation is also called for since
only then one can be confident about the estimated theoretical
uncertainty. Work in this direction is in progress.

\noindent 
{\bf Acknowledgments}

We would like to thank J.~Niskanen for his help to identify the source
of the discrepancy between parts of our calculation and Ref.~\cite{kolck}
and G.A. Miller, U. van Kolck and D.R. Bolton for useful discussions.
Work supported in parts by funds provided from the Helmholtz
Association (grants VH-NG-222, VH-VI-231) and by the DFG (SFB/TR 16
and DFG-RFBR grant 436 RUS 113/991/0-1) and the EU HadronPhysics2
project.  V.B. and A.K. acknowledge the support of the Federal Agency of
Atomic Research of the Russian Federation.

\end{document}